# A Systematic and Analytical Review of the Socioeconomic and Environmental Impact of the Deployed High-Speed Rail (HSR) Systems on the World

[1*]Mohsen Momenitabar, [2]Zhila Dehdari Ebrahimi, [3]Mohammad Arani

[1*]College of Business, Transportation Logistics & Finance, North Dakota State University, Fargo, ND 58108-6050, USA; Email: Mohsen.momenitabar@ndsu.edu; ORCID 0000-0003-2568-1781.

[2]College of Business, Transportation Logistics & Finance, North Dakota State University, Fargo, ND 58108-6050, USA; Email: zhila.dehdari@ndsu.edu; ORCID 0000-0001-7256-0881.

[3]Systems Engineering Department, University of Arkansas at Little Rock, Little Rock, AR, 72204, USA; Email: Mxarani@ualr.edu; ORCID 0000-0002-1712-067x.

*Corresponding Author

**Abstract:**

The installation of high-speed rail in the world during the last two decades resulted in significant socioeconomic and environmental changes. The U.S. has the longest rail network in the world, but the focus is on carrying a wide variety of loads including coal, farm crops, industrial products, commercial goods, and miscellaneous mixed shipments. Freight and passenger services in the U.S. dates to 1970, with both carried out by private railway companies. Railways were the main means of transport between cities from the late 19th century through the middle of the 20th century. However, rapid growth in production and improvements in technologies changed those dynamics. The fierce competition for comfortability and pleasantness in passenger travel and the proliferation of aviation services in the U.S. channeled federal and state budgets towards motor vehicle infrastructure, which brought demand for railroads to a halt in the 1950s. Presently, the U.S. has no high-speed trains, aside from sections of Amtrak's Acela line in the Northeast Corridor that can reach 150 mph for only 34 miles of its 457-mile span. The average speed between New York and Boston is about 65 mph. On the other hand, China has the world's fastest and largest high-speed rail network, with more than 19,000 miles, of which the vast majority was built in the past decade. Japan's bullet trains can reach nearly 200 miles per hour and dates to the 1960s. That system moved more than 9 billion people without a single passenger casualty. In this systematic review, we studied the effect of High-Speed Rail (HSR) on the U.S. and other countries including France, Japan, Germany, Italy, and China in terms of energy consumption, land use, economic development, travel behavior, time use, human health, and quality of life. In order to address the significance of the subject, we have collected more than 90 peer-reviewed articles from scholarly academic databases including Elsevier, Springer, Emerald, and Taylor & Francis. We augmented each article with the authors' analytical points of view, emphasizing and weighting the socioeconomic and environmental factors. From this synthesis and cognitive assessment, we suggest potentially effective approaches and policies to enhance the resilience of the system and to fortify sustainable development. Our studies reveal that HSR deployments had the lowest effect on human health depreciation, elevated quality of life, and maintained sustainability. Additionally, it had an unprecedented impact on economic development, travel behavior, and travel time. Many methods have been employed to analyze these impacts. Methods include regression analysis, Monte Carlo simulation, cost-benefit analysis, prioritization, use of sensor technologies, Tobit regression analysis, spatial analysis, continuous improvement strategy, competition model between airline and high-speed rail, simple supply-oriented regional econometric model, game-theoretic frameworks, dynamic panel modeling, nonlinear regression models, emission inventory models fused with construction material models, diesel and electricity models, emissions factors for construction equipment, heavy-duty trucks, technological optimism, nested logit models, and regional accessibility perspectives. Moreover, a number of papers reported either positive or negative impacts on factors such as human health and lifestyle. The contribution of this systematic review will help readers to grasp the profound impacts of HSR on socioeconomic and environmental characteristics.

## 1. Introduction:

The implementation of high speed rail (HSR) in US will provide Americans people with more transportation choices. In fact, as the passenger grows, the Rail industry need to be updated to transfer more passengers across the US. Additionally, the quality and speed play the key roles in this way. Investment in high speed rail in US can have an advantage point like creating jobs, increasing the economic activities, Boosts Productivity, Reduces the Nation's Dependence on Foreign Oil, Expands Travel Choices, and Improves Mobility. With the extending the Bullet trains fuel real-estate, the quality of life will improve, the air pollution and traffic congestion will reduce, and provide a "safety valve" for crowded cities, especially in the developing world, according to a study by Chinese and U.S. economists. The study is based on China's high speed rail network, but the researchers said the benefits experienced there would be the same for California's proposed high speed rail system. Bullet train systems connecting China's largest cities to nearby smaller cities have made cities more attractive for workers and alleviated traffic congestion and pollution in megacities which carried out by Tsinghua University and the University of California, Los Angeles.

In this paper, we will examine the effect of high speed rail in US and other countries like French, Japan, Germany, Italy, and China on energy consumption, land use, economic development, travel behavior, time use, human health, and quality of life. We focused on different aspects of the interconnected HSR systems to evaluate the impact of it. However, there is a lack of understanding of the extent to which the world's now mature HSR infrastructure on the environmentally and socioeconomically aspect of it. In the remaining sections, we will focus deeply on Interconnected HSR in the world especially in US.

### 1.1 High Speed Rail in the world

High speed rail (HSR) is a sort of rail transport framework that works essentially quicker than conventional rail traffic, utilizing a coordinated arrangement of moving stock and devoted tracks. In fact, it is considered as one of the most dominant mechanical improvement in moving travelers between different pieces of the world. (Javier campos, 2009) Gathered information of 166 high-speed rail in all over the world to analyze the crucial problem for conducting of transportation industry. Firstly, he described the differences between development and operations models. After that, they considered the cost of building, and cost of maintenance of the high speed rail infrastructure. His aim was economic analysis of future projects, better understanding of the expected construction and operations cost, number of passengers, and the growth rate which must be done based on various economic and geographic situation.

The first high speed rail system was created in Japan, between Tokyo and Osaka, in 1964. It was widely known as the bullet train. The Japan has the most mile of rail in the world with 1524 mile, while, the Britain has the less mile of rail with 70 miles. (June, 2009). In 2009, there were 7555-mile high speed rail in all over the world, while in 2018 this reached approximately to 49000. France was the next country to make high speed rail available to the public in 1981, serving between Paris and Lyon at 124 mph. The French high speed rail network now comprises over 2,800 km of Lignes à Grande Vitesse (LGV), which allows speeds of up to 200 mph running on its TGVs.

High speed rail in China has developed rapidly over the past 15 years thanks to generous funding from the Chinese government. In the early 1990s, China started planning the current high speed rail network, modeling it after the Shinkansen system in Japan. In table 1, we can see the length of line in operation, under construction, and not started construction according to their level of deployment of HSR railways, in order from most development to least, based on data from the International Union of Railways (UIC) and from other sources that provide updated data in all over the world.

| **Country** | **Length of lines in operation (km)** | **Lines under construction (km)** | **Approved but not started construction** | **Max speed (km/h)** |
|---|---|---|---|---|
| China | 26,869 | 10,738 | **1,268** | 350 |
| Spain | 3,100 | 1,800 | 0 | 310 |
| Japan | 3,041 | 402 | 194 | 320 |
| France | 3,220 | 125 | 0 | 320 |
| Germany | 3,038 | 330 | 0 | 300 |
| Sweden | 1,706 | 11 | 0 | 205 |
| United Kingdom | 1,377 | 230 | 320 | 300 |
| South Korea | 1,104 | 376 | 49 | 305 |
| Italy | 999 | 116 | 0 | 300 |
| Turkey | 802 | 1,208 | 1,127 | 300 |
| Russia | 845 | 0 | 770 | 205 |
| Finland | 609 | 0 | 0 | 220 |
| Uzbekistan | 600 | 0 | 0 | 250 |
| Austria | 352 | 208 | 0 | 250 |
| Taiwan-China | 354 | 0 | 0 | 300 |
| Belgium | 326 | 0 | 0 | 300 |
| Poland | 224 | 0 | 484 | 200 |
| Netherlands | 175 | 0 | 0 | 300 |
| Switzerland | 144 | 15 | 0 | 250 |
| Luxembourg | 142 | 0 | 0 | 320 |
| Norway | 64 | 54 | 0 | 210 |
| ***U.S.A.*** | 54 | 192 | 1,710 | 240 |
| Saudi Arabia | 0 | 453 | 0 | 300 |
| Denmark | 0 | 56 | 0 | 200 |
| Thailand | 0 | 0 | 615 | 300 |
| Sweden | 0 | 11 | 0 | 205 |
| Russia | 0 | 0 | 770 | 250 |
| Iran | 0 | 0 | 1,351 | 300 |
| Indonesia | 0 | 0 | 712 | 250 |
| India | 0 | 0 | 508 | 250 |
| Malaysia/Singapore | 0 | 0 | 350 | 250 |
| Israel | 0 | 0 | 85 | 250 |
| Portugal | 0 | 0 | 550 | 250 |
| Czech Republic | 0 | 0 | 660 | 250 |
| Greece | 0 | 500 | 200 | 250 |
| Hungary-Romania | 0 | 0 | 460 | 250 |

Table 1. The length of line in operation, under construction, and not started construction based on data from the International Union of Railways (UIC);

Source: UIC, 2010 and 2011

### 1.2 The history of high speed rail in US

High speed rail (HSR) in United States does not attract more passenger due to the various factors. We can answer to this question by presenting this question that why are other countries investing more money on High-Speed Rail? High speed rail should be viewed as a complicated system that incorporates foundation, moving stock, flagging frameworks, upkeep frameworks, stations, activity the executives, financing and legitimate viewpoints, among different segments.

Freight and passenger services in the US goes back to 1970. Both it were carried out by private railway companies. Rail was the main means of transport between the cities from the latter part of the 19th century to the middle of the 20th century, but the dynamics were also modified by increasing production patterns and new techniques. The competition from motor vehicles using the U.S. and intergovernmental road system and an increasing aviation infrastructure pushed rail carriers to ruin in the 1950s. In 1967, New York Centrally, the exclusive 20th century, took its last ride. The National Railroad Passenger Corporation (Amtrak) was founded by the Congress in 1970. AMTRAK is a government-owned company set up to establish a national network for rail passengers. It also allowed private railway companies to transfer to AMTRAK their passenger train operations by allowing them to make money. The most direct benefit of the AMTRAK is that it would provide the opportunity for long, intermediate, and relatively short-distance trips, serving a wide range of travelers, whether for business, daily commuting, or leisure. The high-speed train would be a strong viable transportation alternative for relatively longer distance travel as door-to-door travel times would be comparable to air travel and less than one-half as long as an automobile trip.

In the terms of the environmental benefits of AMTRAK, electrically powered high-speed trains would reduce reliance on gasoline consumption. In addition to environmental benefits, the AMTRAK system as a whole would produce important spillover benefits or positive externalities that never anticipate riding on a High Speed train will enjoy. Moreover, the AMTRAK network is expected to provide grade separations, so traffic delays at existing at grade crossings will be diminished to the extent that it separates the grades of all tracks where the AMTRAK system shares rights-of-way. Perhaps the most significant positive externality of AMTRAK is the expanded economic activity resulting from lowering the transportation costs of a region and expanding its accessibility to broader product and labor markets.

The main challenges of high speed rail in US including the application to the transport geography of the US, stop access, and right-of-way acquisition. Actually, the train should be able to compete with air travel and automobile by improving their speed as compared to the airplane and automobile (Lane, 2012).

The below shape has shown the length of line in operation, under construction, and not started construction.

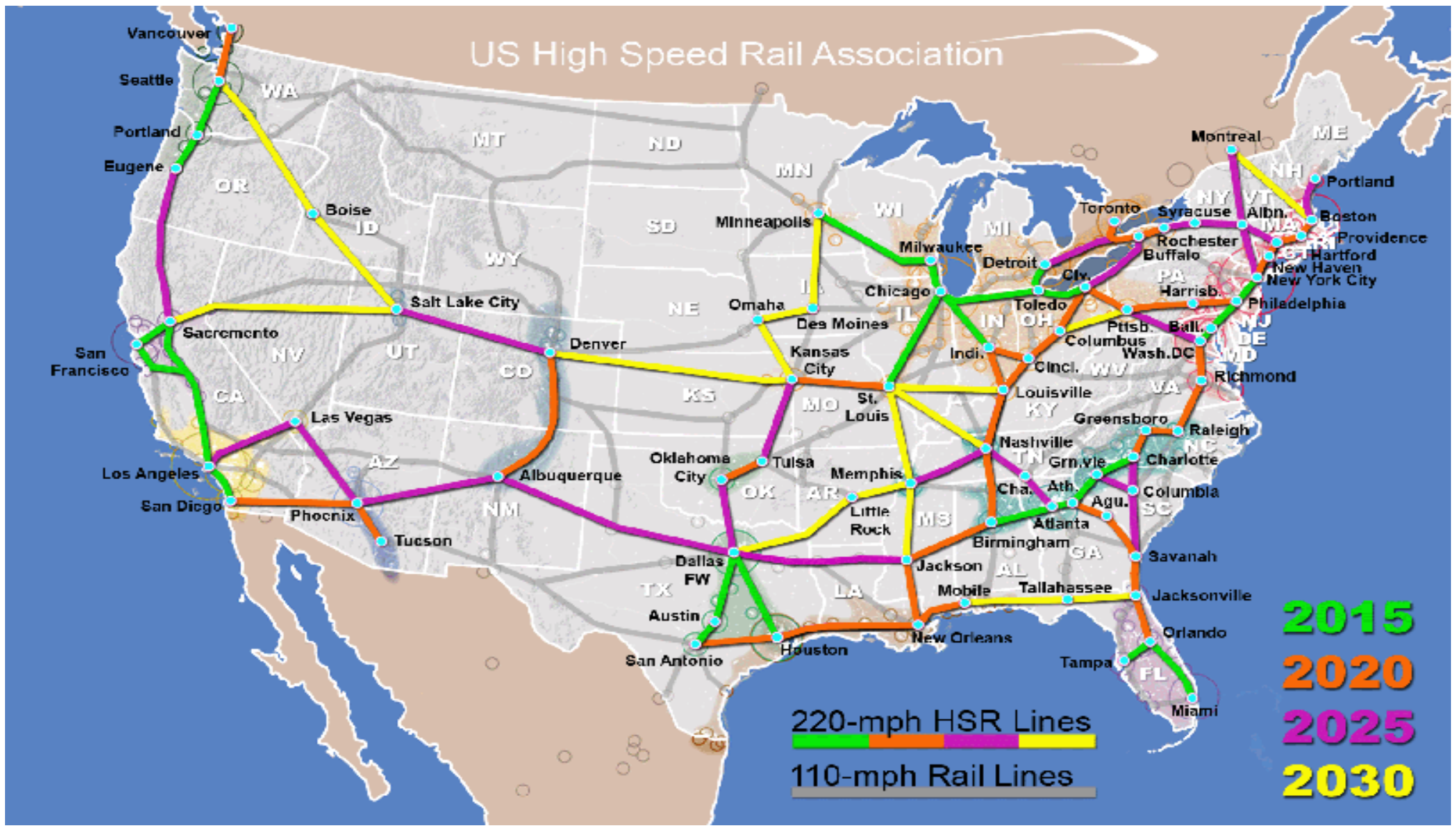

Fig 1. The length of line in operation, under construction, and not started construction based on data from the International Union of Railways (UIC); Source: UIC, 2010 and 2011

## 2. Literature Review:

Today, most of countries such as France, Germany, Italy, Spain, Japan and China already developed the interconnected High speed rail systems at various scale. On the other hand, other countries such as the United States, India and Malaysia, are still debating whether the massive costs of HSR infrastructure development could be offset by the benefits that the system is expected to generate. Many studies have tried to evaluate the socioeconomic impact of the deployed HSR systems. Furthermore, the transportation system is facing considerable challenges to manage its environmental impacts to minimize disruptions to human society and the climate. The increasing number of destabilizing events, such as extreme weather events, man-made hazards, and technological system failures, have raised concerns about the sustainable operation of HSR systems over the long run.

In table 2, the impact of High speed rail on energy consumption, land use, economic development, travel behavior, time use, human health, quality of life, etc. has been summarized which is done by various researchers and published in different peer-reviewed journals.

Table2: The impact of High speed rail on energy consumption, land use, economic development, travel behavior, time use, human health, and quality of life.

| **Author** | **Energy Consumption and Climate Change** | **Emissions and Air Quality(environmental impact)** | **Land use including Sprawl, Community Severance, and Risks to Biodiversity** | **Economic Development, Travel Behavior, and Time Use** | **Human Health and Quality of Life** | **Assessment of the Vulnerability and Resilience** | **Policy Analysis to Support Developing a Resilient** | **New Approaches** |
|---|---|---|---|---|---|---|---|---|
| (Hang Yuan, 2019) | | | | | ✓ | | | |
| (Javier campos, 2009) | | | | ✓ | | ✓ | | |
| (Regina R. Clewlowa, 2014) | | ✓ | | ✓ | ✓ | | | |
| (kamga , 2015) | | | | ✓ | | ✓ | ✓ | |
| (Marie Delaplace , 2015) | | | | ✓ | | | | |
| (Óscar Álvarez-SanJaime, 2016) | | | | ✓ | | | ✓ | |
| (Wangtu Xu, 2019) | | | ✓ | ✓ | | | | |
| (Paul Chinowsky , 2019) | ✓ | | | | | ✓ | | |
| (Vickerman , 2018) | | | | ✓ | | | | |
| (Amparo Moyano, 2018) | | | | ✓ | | ✓ | | |
| (Hyojin Kim, 2018) | | | ✓ | ✓ | | | | |

| (Tiziana D’ Alfonso, 2015) | (Venkat Krishnan, 2015) | (Andrés Monzón, 2013) | (De-gen Wang, 2018) | (Begoña Guirao, 2017) | (Maddi Garmendia, 2012) | (Nikhil Bugalia, 2019) | (Longfei Zheng, 2019) | (Xiaoyan Chen, 2020) | (Rui Zhang, 2019) | (Anming Zhang, 2019) | (Hongchang Li, 2019) | (Wang, 2018) | (Paolo Beria, 2018) |
|---|---|---|---|---|---|---|---|---|---|---|---|---|---|
| | ✓ | | | | | | | | | | | | |
| | | | | | | | | | | ✓ | | | |
| | | ✓ | ✓ | | ✓ | | ✓ | ✓ | | | | ✓ | |
| ✓ | ✓ | | ✓ | | | | | | ✓ | | ✓ | | ✓ |
| ✓ | | | | | | | | | | ✓ | | | |
| | | ✓ | | | | ✓ | | | ✓ | | | | |
| | | | | | ✓ | | | | ✓ | | | | |
| | ✓ | | | ✓ | | | ✓ | | | | ✓ | | ✓ |

| (Li, 2019) | (Schipper, 2011) | (Verma, 2019) | (Komei Sasaki, 1997) | (Heather Jones, 2017) | (JIN Fengjun, 2017) | (Hiramatsu, 2018) | (Corinne Blanquart, 2017) | (Pierre Zembri, 2017) | (E.Doomernik, 2015) | (Cécile Chèze, 2017) | (Varun raturi, 2017) |
|---|---|---|---|---|---|---|---|---|---|---|---|
| | ✓ | | | ✓ | | | | | | | |
| | | | | | | | | | | | |
| | | | | | | | | | | | |
| ✓ | | | ✓ | | | ✓ | ✓ | ✓ | | ✓ | ✓ |
| | | | | | | | | | | | |
| | | | | | | | | ✓ | | | |
| | | | | | | | | ✓ | | | |
| | | ✓ | | | ✓ | ✓ | | ✓ | ✓ | ✓ | |

| (Juan M.Matute, 2015) | (Camille Kamga, 2014) | (Bin Xu, 2013) | (Jonas Westin, 2012) | (Brenda Chang, 2011) | (Åkerman , 2011) | (Bert van Wee, 2003) | (Bo Wang, 2019) | (Fan Zhang, 2019) | (Ya-Yen Sun, 2018) | (Haynes , 2015) | (Shi, 2019) | (Long Ye, 2019) |
|---|---|---|---|---|---|---|---|---|---|---|---|---|
| ✓ | ✓ | ✓ | ✓ | ✓ | ✓ | | | | | | | |
| ✓ | ✓ | ✓ | ✓ | ✓ | | ✓ | | | | | | |
| | | | | | | | | | | | | |
| ✓ | | | | | | | ✓ | ✓ | ✓ | | | |
| | | | | | | | | | | | | ✓ |
| | ✓ | | | | | | | | | ✓ | | |
| | | | | | | | | | | | | |
| | | | | | | | | | | ✓ | ✓ | |

| (Shih-Lung Shaw, 2014) | (Héctor S. Martínez Sánchez-Mateos, 2012) | (Ennio Cascetta, 2011) | (Chen, 2017) | (Juan Luis Campa, 2016) | (Hyojin Kim, 2015) | (Jing Cao, 2013) | (Zhenhua Chen, 2019) | (Bruno Dalla Chiara, 2017) | (Robertson, 2016) | (Guizhen He, 2016) | (Ye Yue, 2015) |
|---|---|---|---|---|---|---|---|---|---|---|---|
| | | | | | | | ✓ | ✓ | | | |
| | | | | | | | | | ✓ | | ✓ |
| ✓ | | | | | | | | | | | |
| ✓ | ✓ | ✓ | | ✓ | ✓ | ✓ | | ✓ | | ✓ | |
| | | | | | | | | | | | |
| | | | | | | | ✓ | | | | ✓ |
| | | | | | | | | ✓ | | ✓ | |
| | | | ✓ | | | | | | ✓ | | |

| (Qiong Zhang, 2017) | (Jian Zhao, 2015) | (Tao Li, 2019) | (Diao, 2018) | (Tsunoda, 2018) | (Levinson, 2012) | (Ping Yin, 2019) | (Woodburn , 2019) | (Miao Yu, 2018) | (Xijing Li, 2016) | (Zhenhua Chen, 2017) | (Jingjuan Jiao, 2017) | (Lvhua Wang, 2016) | (Yu Shen, 2014) |
|---|---|---|---|---|---|---|---|---|---|---|---|---|---|
| | | | | | | | | | | | | | |
| | | | | | | | | | | | | | |
| | | | ✓ | | | | | | | ✓ | ✓ | | ✓ |
| ✓ | ✓ | | | | ✓ | ✓ | | ✓ | ✓ | | | ✓ | |
| | | | | | ✓ | | | | | | | | |
| | | | | | | | | | | | | | |
| | | ✓ | | | | | ✓ | | | | | | |
| | | | | ✓ | | | | ✓ | ✓ | | ✓ | ✓ | ✓ |

| (Daniel Albalate, 2016) | (ShuliLiu, 2019) | (Guizhen He, 2015) | (Gulcin Dalkic, 2017) | (Changmin Jiang, 2016) | (Yannick Cornet, 2018) | (Zhe Chen, 2019) | (Feng Wang, 2019) | (Shuli Liu, 2019) | (Wangtu(Ato) Xu, 2018) | (Fangni Zhang, 2018) | (Shanming Jia, 2017) | (Zhenhua Chen, 2016) |
|---|---|---|---|---|---|---|---|---|---|---|---|---|
| | | | ✓ | | | | | | | | | |
| | | ✓ | ✓ | | ✓ | | | | | | | |
| | ✓ | | | | | | ✓ | ✓ | | | | |
| ✓ | | | | ✓ | | | ✓ | | | | ✓ | ✓ |
| | | | | | | | | | | | | |
| | | | | | | | | | ✓ | ✓ | | |
| | | | | | | | | | | | | |
| | | | | | | ✓ | | | | ✓ | ✓ | |

| (Xuezhen Dai, 2018) | (Benjamin R.Sperry, 2017) | (Yanyan Gao, 2020) | (SONG Xiao-dong, 2014) | (Lvhua Wang, 2016) | (Cheng Jin, 2013) | (Silva, 2013) | (Chen, 2019) | (Degen Wang, 2014) | (Jin Weng, 2020) | (Gualter Couto, 2015) | (Dongrun Liu, 2019) | (Francesca Pagliara, 2020) |
|---|---|---|---|---|---|---|---|---|---|---|---|---|
| | | | | | | | | | | | | |
| | | | ✓ | | | ✓ | | | | | | |
| | | | | | ✓ | | | | | | | ✓ |
| ✓ | | | | ✓ | | | ✓ | ✓ | ✓ | ✓ | | |
| | | | | | | | | | | | | |
| | | | | | | | | | | | ✓ | |
| | | | | | | | | | | | | |
| | ✓ | ✓ | | | | | | | | | | |

| | | | | | | | | |
|---|---|---|---|---|---|---|---|---|
| (Petra Kaczensky, 2003) | | | ✓ | | | | | |
| (Begoña Guirao, 2017) | | | | ✓ | | | | |
| (Céline Clauzel, 2013) | | | | ✓ | | | | |
| (Carlos Llorca, 2018) | | | | ✓ | | | | |
| (Chunyang Wang, 2020) | | | | ✓ | | | | |
| (Qiong Zhang, 2020) | | | | ✓ | | | | |
| (Sergej Bukovac, 2019) | | ✓ | | | | | | |
| (Min Su, 2019) | | | | ✓ | | | | |
| (Guineng Chena, 2014) | | | | ✓ | | | | |
| (Robusté, 2005) | | | | ✓ | | | | |
| (Emilio Ortega, 2012) | | | ✓ | ✓ | | | | |
| (Zhenhua Chen, 2015) | | | | ✓ | | | | |
| (Liwen Liu, 2018) | | | | ✓ | | | | |
| (Mohammad Arani, 2020) | | | | ✓ | | | ✓ | |

In fact, most of the peer-review article considered on the table 2 have gotten from the database like Elsevier, springer, and Taylor Francis. In table 2, we have collected already 116 articles and categorized the effect of HSR on the environment and economics so that effective planning and policy strategies could be developed to enhance the resilience of the system and support the goal of sustainable development.

Form the table 2, we can deduce that the interconnected HSR have a highest impact on economic development, travel behavior, time use, and new approaches. While, the Interconnected HSR systems had a lowest effect on human health, quality of life and policy analysis to support developing a resilient.

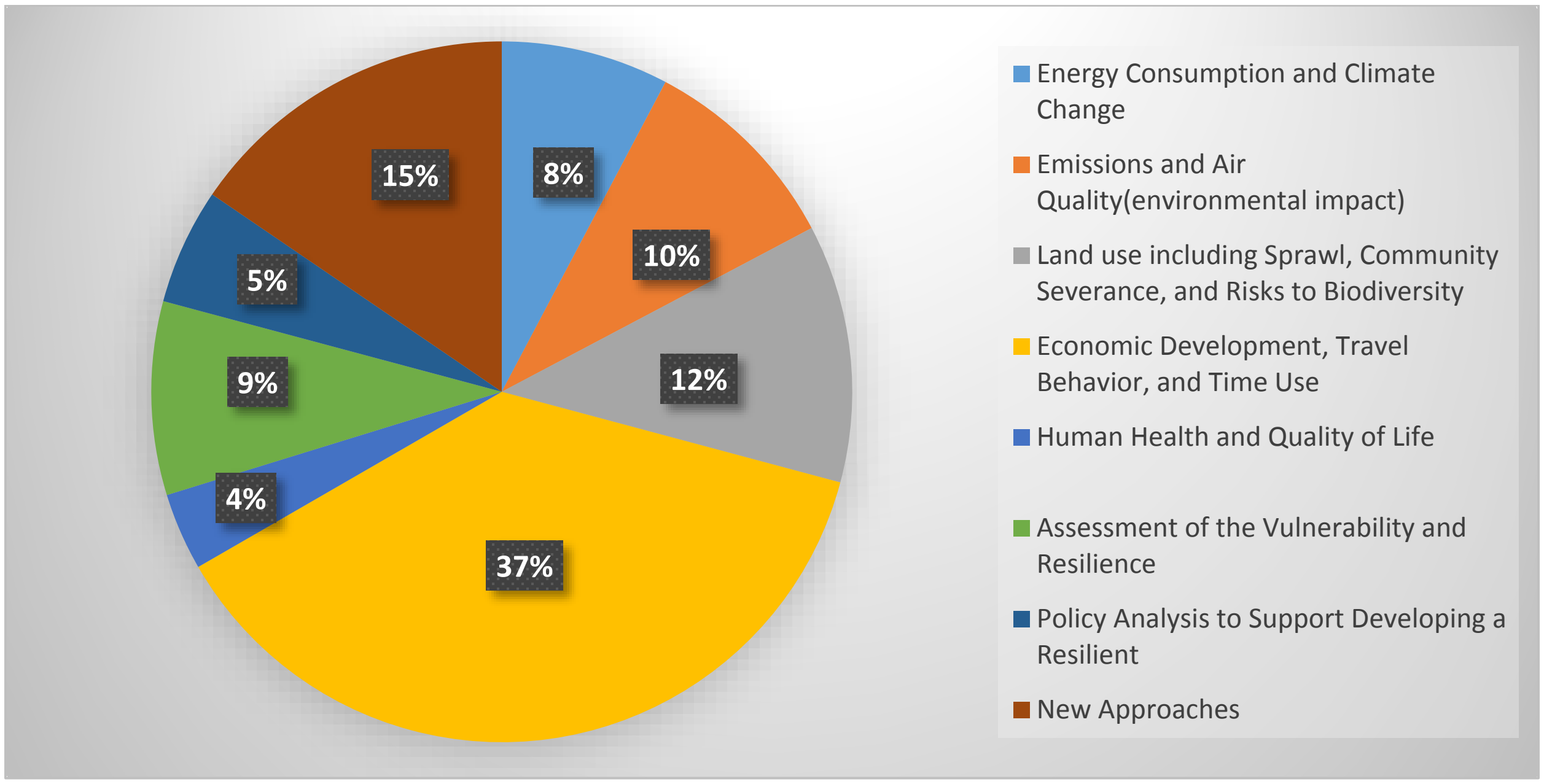

Figure 2: Number of articles which have effected on different criteria

In figure2, we can deduce that high speed rail has the most effect on economic development, travel behaviors and time use. On the other hand, the high speed rail has the lowest effect on human health and quality of life.

## 3. Methodology

In this section, we will discuss the methodology which our selected publications investigated in their works. We have summarized the information in table 3, based on different categories like what measurement the author applied, the intensity level of impact (poor, moderate, good), and finally what method to conclude the impact.

Table3: Methodology

| **Author** | **What measurement** | **What method to conclude the impact** |
|---|---|---|
| (Hang Yuan, 2019) | The reform's effect on import by rail | Difference in-differences (DID) regressions |
| (Javier campos, 2009) | Actual cost values of building and maintaining a high speed rail infrastructure | The economic definition of high-speed rail |
| (Regina R. Clewlowa, 2014) | Demand | Regression analysis |
| (kamga, 2015) | Station hubs | Urbanization of three mode of passengers including road, rail, air |
| (Marie Delaplace, 2015) | | Analysis fare is done by simulating internet-based bookings. |
| (Óscar Álvarez-SanJaime, 2016) | Vertical structure of the rail sector | Simulation |
| (Wangtu Xu, 2019) | Regional structure factors, coverage factors, and urban land development potential | Prioritization |
| (Paul Chinowsky, 2019) | Delay costs | Use of sensor technology combined with changes to operating procedures |
| (Vickerman, 2018) | Demand and cost | Numerical simulation |
| (Amparo Moyano, 2018) | Potential connectivity between stations and the other based on actual HSR services | The schedule-based approach, The location-based approach |
| (Hyojin Kim, 2018) | Locations | Node-place model for classifying the performance of station areas |
| (Paolo Beria, 2018) | Demand and cost | Planning optimism |
| (Wang, 2018) | Regional accessibility | Tobit regression analysis |
| (Hongchang Li, 2019) | Demand | Difference-in-differences approach |
| (Anming Zhang, 2019) | Competition, speed, market power, cooperation | Network feature |
| (Rui Zhang, 2019) | Fares and frequency | Ex-post analysis |
| (Xiaoyan Chen, 2020) | Three levels of authors, institutions, and countries | Social network analyses and graph theory |
| (Longfei Zheng, 2019) | Spatial spillover effect of HSR stations in China | Difference-in-differences approach |
| (Nikhil Bugalia, 2019) | The risk and control sharing | Continuous improvement strategy |
| (Maddi Garmendia, 2012) | - | - |
| (Begoña Guirao, 2017) | HSR as a variable and inter-regional labor mobility indicators | Regression analysis |
| (De-gen Wang, 2018) | Urban tourism and nationwide accessibility of cities in non-HSR and HSR networks | Spatial analysis in ArcGIS |
| (Andrés Monzón, 2013) | Major urban areas, distribution of accessibility values | Spatial impact analysis techniques based on the computation of accessibility indicators, supported by a Geographical Information System (GIS) |
| (Venkat Krishnan, 2015) | Energy flow, investment in generation technology, number of trips, and yearly fleet investments in transportation mode | Cost-minimization network equilibrium model |
| (Tiziana D'Alfonso, 2015) | Frequency and train speed | Competition model between one airline and high speed rail over a single origin (O) – destination (D) link |
| (Varun raturi, 2017) | Demand | Nash equilibrium of the game method |
| (Cécile Chèze, 2017) | Labor and intermediary goods and services markets | Cost-benefit analysis |
| (E.Doomernik, 2015) | Network and the rolling stock assets | Network Data Envelopment Analysis (NDEA) in combination with the Malmquist Productivity Index |
| (Pierre Zembri, 2017) | Territorial coverage | Methods of planning and funding |
| (Hiramatsu, 2018) | Context of regional economies and transportation | Computable general equilibrium analysis |
| (JIN Fengjun, 2017) | Transport circle and accessibility of HSR in East Asia | "Core-core" model |
| (Heather Jones, 2017) | Train manufacturing, train operation, train maintenance, train disposal, track construction, track operation, and maintenance, and track disposal | The SimaPro Life Cycle Assessment software |
| (Komei Sasaki, 1997) | Economic activities and population | Simple supply-oriented regional econometric model |
| (Verma, 2019) | Fare and frequency | A game-theoretic framework |
| (Schipper, 2011) | Energy use and greenhouse gas emissions | Descriptive analysis of cross-national passenger transport trends |
| (Li, 2019) | Tourism economy | Propensity score matching and difference-in-difference methods |
| (Long Ye, 2019) | Behaviors of high speed rail drivers | Questionnaire method includes five parts: instructing, big five personality, organizational identification and safety performance. |
| (Shi, 2019) | Tourism arrivals | Nonlinear regression model |
| (Haynes, 2015) | Tourism demand | Dynamic panel modeling |
| (Ya-Yen Sun, 2018) | Travel patterns of individual tourists | Questionnaire survey |
| (Fan Zhang, 2019) | Investment | Difference-in-difference and propensity score matching tests |
| (Bo Wang, 2019) | Nature of ICT device use and in-vehicle e-activities | Questionnaire survey |
| (Bert van Wee, 2003) | Rail and other major infrastructure projects | Cost benefit analysis |
| (Åkerman, 2011) | Life cycle | Combination of simplified life-cycle assessment (LCA) and scenarios including future developments regarding vehicle technology, biofuels, mix of electricity supply and transport volumes |
| (Brenda Chang, 2011) | High speed rail infrastructure | The emission inventory model combines data from construction material LCAs, diesel and electricity LCAs, and emissions factors for construction equipment and heavy duty trucks |
| (Jonas Westin, 2012) | Future transport demand, technology and power production | Monte Carlo simulation |
| (Bin Xu, 2013) | High speed rail carriages | Ten measurement points at supply air outlet, passenger-breathing zone and recirculation air vents are used. |
| (Camille Kamga, 2014) | High costs of “car culture” and air travel in the U.S. | Technological optimism |
| (Juan M. Matute, 2015) | Capital costs fail to capture indirect benefits | Cost-benefit methodology |
| (Ye Yue, 2015) | Vehicle, infrastructure, operation | Different capacity utilization rate scenarios |
| (Guizhen He, 2016) | Public protests | Decision-making |
| (Robertson, 2016) | Mitigation of CO2 emissions | LCI (life cycle inventory) |
| (Bruno Dalla Chiara, 2017) | Travel time and distance | Simulation tools |
| (Zhenhua Chen, 2019) | HSR and aviation | Artificial intelligence technique including data visualization and regression analysis |
| (Jing Cao, 2013) | China’s HSR network | Weighted average travel times and travel costs, contour measures, and potential accessibility are employed as indicators of accessibility analysis |
| (Hyojin Kim, 2015) | The accessibility of each stage of HSR network | Weighted average travel time and potential accessibility indicators |
| (Juan Luis Campa, 2016) | Tourism | Multivariate panel analysis |
| (Chen, 2017) | Domestic air transportation | Improved panel regression model by taking into account of the detailed opening schedules of various HSR services |
| (Ennio Cascetta, 2011) | Demand | Nested Logit model |
| (Héctor S. Martínez Sánchez-Mateos, 2012) | Accessibility of a station | Regional accessibility perspective |
| (Shih-Lung Shaw, 2014) | Travel time, travel distance and ticket fee. | Timetable-based accessibility evaluation approach |
| (Yu Shen, 2014) | Land cover change | Spatial mixed logit methods with panel data structure |
| (Lvhua Wang, 2016) | Travel time and accessibility | A layer cost distance (LCD) method and A door-to-door approach |
| (Jingjuan Jiao, 2017) | Passenger | Graph index, average path length, clustering coefficient and three indicators of centrality, namely weighted degree centrality (WDC), weighted closeness centrality (WCC) and weighted between ness centrality (WBC). |
| (Haynes, 2015) | Regional economic disparity | Three accessibility indicators (weighted average travel time, potential accessibility and daily accessibility) |
| (Xijing Li, 2016) | Redistribution of economic activities | Weighted regression model, geographically network weighted regression (GNWR) |
| (Miao Yu, 2018) | Accessibility scores of the future HSR corridor during peak and off-peak hours | A door-to-door approach |
| (Woodburn, 2019) | Service provision, journey times and train punctuality | Combination of sources including an annual rail freight database, open access real-time train running data, observation surveys and stakeholder interviews. |
| (Ping Yin, 2019) | Tourism spatial interactions | A method of derivation was developed and several indexes, such as tourism mean center, and tourism standard distance |
| (Levinson, 2012) | State of HSR planning | Network structure |
| (Tsunoda, 2018) | HSR and airline | Game-theoretic model |
| (Diao, 2018) | The economic geography | Difference-in-differences analysis and an instrumental variable strategy to address the non-random placement of HSR stations |
| (Tao Li, 2019) | High speed rail (HSR), air transport (AT), conventional railway, expressway and waterway in a state | Complex network theory |
| (Jian Zhao, 2015) | Travel time saving | Time allocation model and the mechanisms of variation in VTTS |
| (Qiong Zhang, 2017) | Demand | Paired T-Test |
| (Zhenhua Chen, 2016) | Economy and environment | Computable general equilibrium model |
| (Shanming Jia, 2017) | Transportation development and economic growth path | DID and PSM-DID method |
| (Fangni Zhang, 2018) | Air travel demand | Difference-in-differences (DID) method |
| (Wangtu(Ato) Xu, 2018) | Connectivity and accessibility indices | The differences in the normalized connectivity-accessibility levels of different categories of cities |
| (Shuli Liu, 2019) | Airport traffic and traffic distribution | Regression analysis |
| (Feng Wang, 2019) | Population mobility and urbanization | Difference-in-difference model |
| (Zhe Chen, 2019) | Air travel | Regression analysis |
| (Yannick Cornet, 2018) | Environmental consequences of major infrastructure decisions | Modal shift |
| (Changmin Jiang, 2016) | Layer geographical coverage and the more polycentric urban system | Comparative research analysis |
| (Gulcin Dalkic, 2017) | Emission of $CO_2$ | Line-based methodology |
| (Guizhen He, 2015) | Transparency, communication and public participation | A series of in-depth interviews with officials of local governments, local Environmental Protection Bureaus (EPBs), and Transportation Bureaus were held; and a Survey among 900 residents along the Beijing-Shanghai HSR |
| (ShuliLiu, 2019) | Centrality (to reflect connectivity) and the harmonic centrality (to reflect accessibility) | Regression analysis |
| (Daniel Albalate, 2016) | Tourism outcomes | Differences-in-differences panel data method |
| (Francesca Pagliara, 2020) | Tourism market | Geographically Weighted Regression technique |
| (Dongrun Liu, 2019) | Driving safety and riding comfort | Multi-body simulation model of a CRH2 high-speed rail vehicle |
| (Gualter Couto, 2015) | Demand threshold | Poisson process |
| (Jin Weng, 2020) | Accessibility | ArcGIS network analysis toolkit |
| (Degen Wang, 2014) | Volumes, flows, and spatial patterns of traffic | Structure of tourism flow network |
| (Chen, 2019) | Regional economic impact | Dynamic and spatial computable general equilibrium-modelling framework |
| (Silva, 2013) | Impacts of general transport infrastructures | Proposes a conceptual model, which is able to combine the different aspects of regional development and jointly evaluate the overall impacts of HSR at the regional level. |
| (Cheng Jin, 2013) | Regional accessibility | Geography information system (GIS) |
| (Lvhua Wang, 2016) | Journey times | Door-to-door approach |
| (SONG Xiao-dong, 2014) | Environmental impact | Life cycle theory, fuzzy analytic hierarchy method, extension theory |
| (Yanyan Gao, 2020) | Innovation | Differences-in-differences method |
| (Benjamin R.Sperry, 2017) | mode choice models | Stated preference (SP) |
| (Xuezhen Dai, 2018) | Surrounding subdivided industries | Agglomeration-diffusion theories |
| (Petra Kaczensky, 2003) | Brown bears | Spatial distribution of bear–vehicle accidents |
| (Begoña Guirao, 2017) | Labor migration | Regression analysis |
| (Céline Clauzel, 2013) | Species distribution | Graph-based approach |
| (Carlos Llorca, 2018) | The economic, social and environmental impacts of transportation | Microscopic long-distance travel demand model |
| (Chunyang Wang, 2020) | Urban economy | The difference-in-difference model with matching method |
| (Qiong Zhang, 2020) | Airline market | Unweighted and weighted Lerner indexes |
| (Sergej Bukovac, 2019) | Australian aviation | Using the existing air travel data, market growth forecast for 2030 on the SYD-CBR-MEL routes, The European competitive envelope for market share capture, The profitability impact for the airlines operating the SYD-CBR-MEL sectors, Australian market variations from European market conditions |
| (Min Su, 2019) | Airlines' intertemporal and average prices | Leisure index, variation coefficient, average discount, and random effects panel data models |
| (Guineng Chena, 2014) | Provincial economic development | Panel Structural Equation Modeling (SEM) formulation |
| (Robusté, 2005) | Air traffic demand | Frequency of service |
| (Mohammad Arani, 2020) | Travel Time | WSDOT algorithm, Wait-Time Search in Decreasing Order of Time |
| (Mohammad Arani, 2019) | Travel Time and Battery Level | FIFO property |

The data which can be summarized form the table 3 are the methods, factors which need to be measured, and quantifying and qualifying the model. Additionally, we have seen that the difference in difference (DID) method, regression analysis, a door-to-door approach, and cost benefit analysis are the popular method which it have been widely used by the authors.

In the figure 3, we see that how many paper which have been analyzed have a positive and negative impacts.

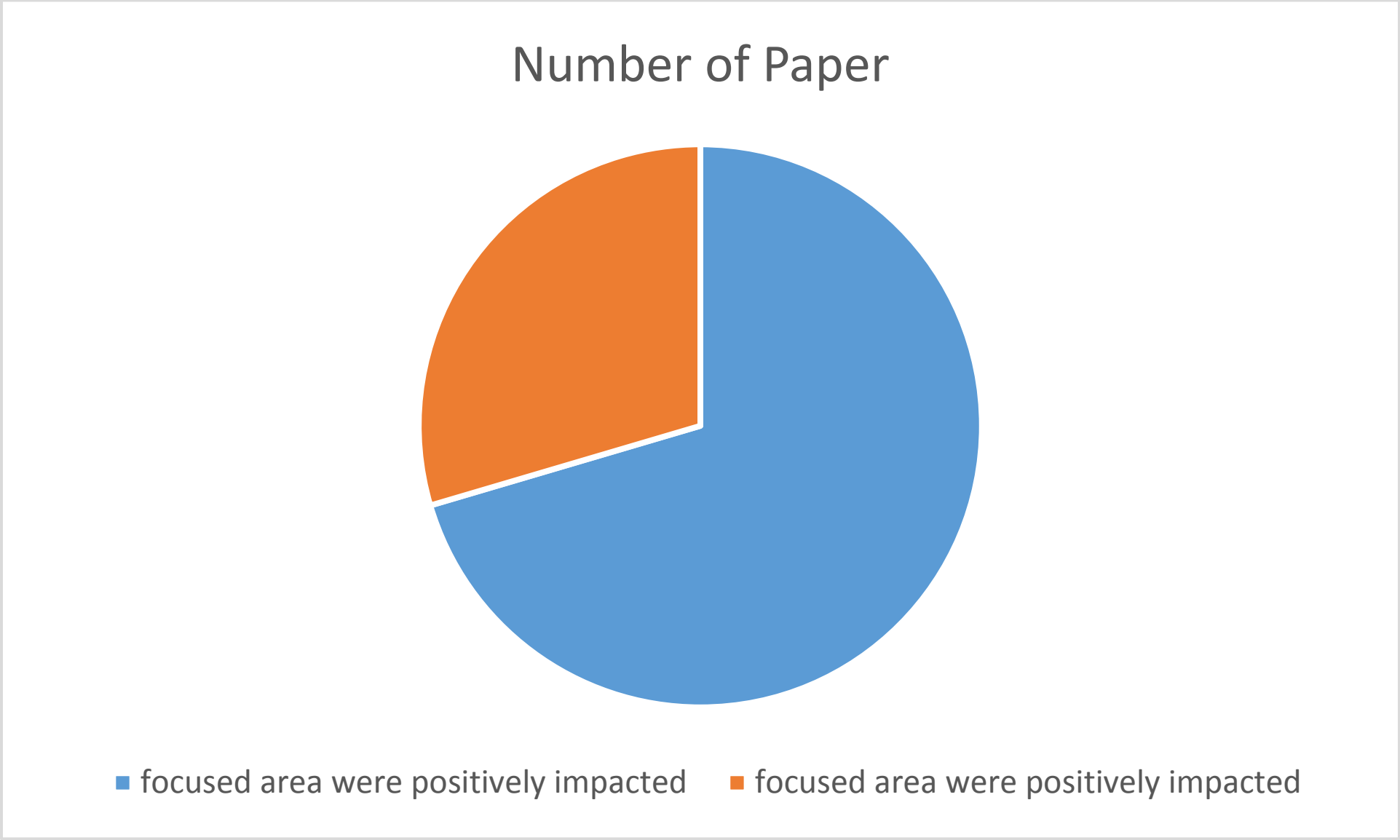

Figure 3: how many paper have a positive and negative impacts

The important point is that, already near of the 65 percent of reviewed paper have the positive effect on HSR and they used to quantified method for explaining their impacts.

In the figure 4, we have summarized all the article which used to in this paper in literature review. I have categorized all of 115 articles which related to the defined categories from year 1997 until 2020.

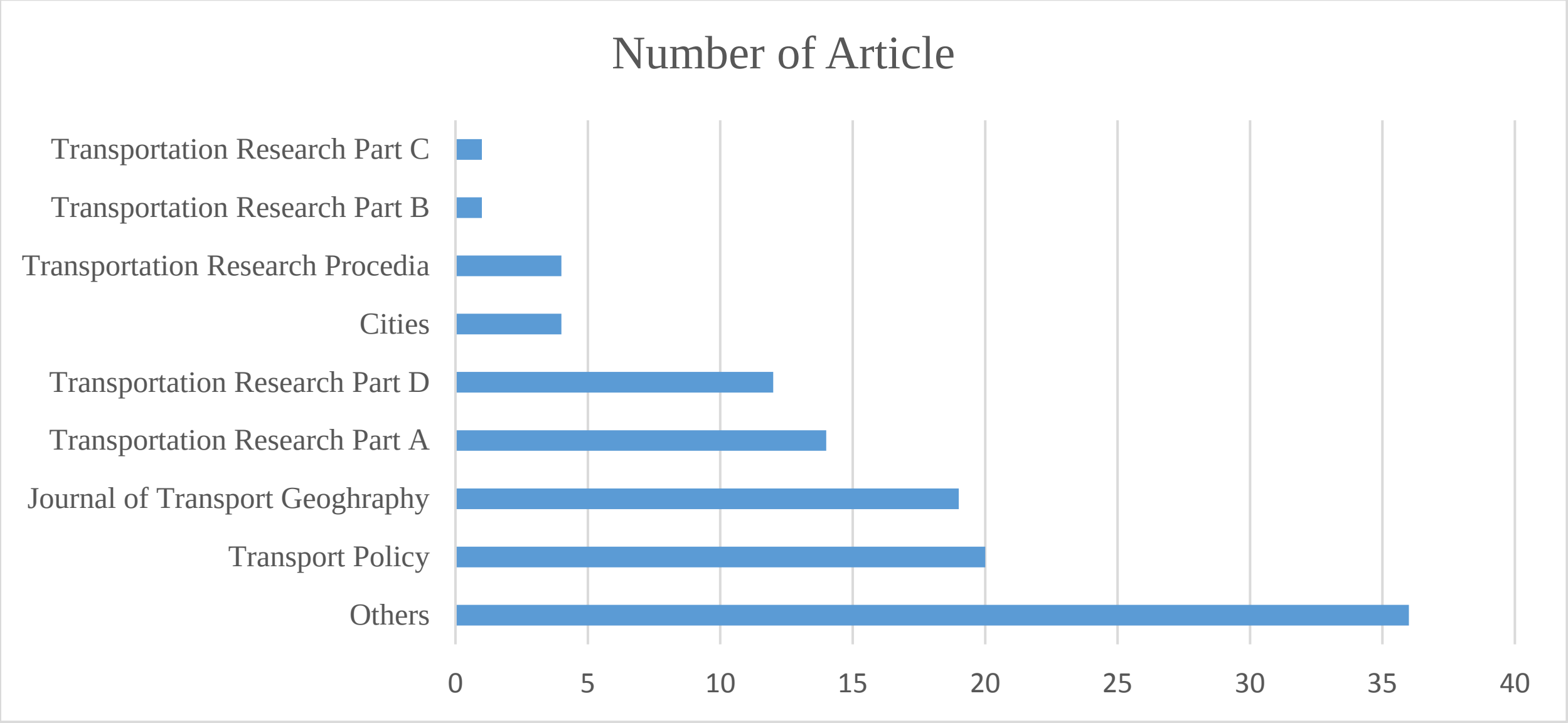

Fig 4. Number of articles used to in literature review

Figure 4 has shown that most of research has been published by the journal of *Transport Policy* and *Journal of Transport Geography, Transportation Research Part A,* and, *Transportation Research Part D.*